# Large CaFeAsF single crystals of high quality grown by the oscillating temperature technique

Ming-Wei MA[a,b†], Binbin Ruan[a,c], Menghu Zhou[a,c], Yadong Gu[a,c], Qingsong Yang[a,c], Junnan Sun[a,c] and Zhi-An Ren[a,c]

[a]*Beijing National Laboratory for Condensed Matter Physics, Institute of Physics, Chinese Academy of Sciences, Beijing 100190, China*

[b]*Songshan Lake Materials Laboratory, Dongguan, Guangdong 523808, China*

[c]*University of Chinese Academy of Sciences, Beijing 100049, China*

**Abstract**

The oscillating temperature technique was successfully employed to grow large and composition homogeneous CaFeAsF single crystals due to the reduction of nuclei number by temperature oscillation. The largest as-grown CaFeAsF single crystal was nearly 8 mm in lateral size and 0.27 mm in thickness which is far larger than previously reported in all dimensions. The full-width-at-half maximum (FWHM) of x-ray rocking curve is as small as 0.163 ° indicating the high quality of CaFeAsF single crystals.

**Keywords:** iron-based superconductivity, crystal growth, oscillating temperature technique, x-ray rocking curve, CaFeAsF

**PACS:** 81.10.Dn, 74.70.−b, 61.05.cp, 68.37.Hk

## 1. Introduction

Since the discovery of superconductivity in F-doped LaFeAsO with $T_C$ = 26 K [1], a series of iron-based parent compounds have been discovered. As one typical member, CaFeAsF with the ZrCuSiAs structure is an analogue of LaFeAsO where the (LaO)+ layer is replaced by the (CaF)+ layer, exhibiting a structural and an antiferromagnetic transition with $T_s$ = 121 K and $T_N$ = 110 K respectively [2-3]. Also, its superconductivity can be tuned by elements substitution or external pressure [2, 4-5] and its $T_C$ can reach 56 K by doping Nd into CaFeAsF [6], which is as high as the $T_C$ record of Sm[O$_{1-x}$F$_x$]FeAs [7]. Moreover, a renewed interest has been triggered by the recent development on the observation of Dirac Fermions in CaFeAsF by quantum oscillation and infrared spectrum [8-9]. The intriguing $T_C$, similar layered structure and distinct physical properties make CaFeAsF an important prototype for studying the mechanism of iron-based superconductivity.

[†] Corresponding author at: Beijing National Laboratory for Condensed Matter Physics, Institute of Physics, Chinese Academy of Sciences, Beijing 100190, China. E-mail: mw_ma@iphy.ac.cn

Single crystal samples of high quality can exhibit truly intrinsic property and anisotropy of the CaFeAsF system, they are therefore in high demand for the investigation of superconducting mechanisms. Many attempts have been made to improve the quality and size of CaFeAsF single crystals grown from CaAs and NaCl flux [3-4, 10-11]. However, up to now, the small crystal size with size of 2 mm is still an obstacle to the further investigation by some state-of-the-art measurement techniques, especially for the inelastic neutron scattering (INS). An INS study on 9 g powder sample of CaFe$_{1-x}$Co$_x$AsF shows the evidence of spin resonance which is capable of providing valuable insight to the superconducting order parameter but lack of detailed momentum information [12]. Therefore, large single crystals are highly demanded by inelastic neutron scattering. We report in this paper that the oscillating temperature technique was successfully employed to grow large CaFeAsF single crystals, which reduces the number of nuclei of CaFeAsF in CaAs flux. The largest as-grown CaFeAsF single crystal was nearly 8 mm in lateral size and 0.27 mm in thickness which is far larger than previously reported in all dimensions. The elements are distributed homogeneously on the as-grown crystal surface with a perfect (0 0 1) orientation shown by the EDX analysis and single crystal XRD. The full-width-at-half maximum (FWHM) fitted from x-ray rocking curve of (0 0 5) Bragg reflection is as small as 0.163° indicating the high quality of CaFeAsF single crystals

2. **Experiments**

Here the principle of this growth technique is sketched as follows. First of all, a careful consideration on the previous crystal growth experiments of CaFeAsF is beneficial for the exploration to the optimal conditions of the oscillating temperature technique. CaFeAsF single crystal with maximum lateral size up to 2 mm was first successfully grown from CaAs flux by a slow cooling with 3 ℃/h from 1150 ℃ to 900 ℃ [10]. In order to improve the growth conditions, the efforts to tune the molar ratio between the self flux CaAs and CaFeAsF, increase the soaking temperature to 1230 ℃ and decrease the cooling rate to 2 ℃/h yield crystals with size still about 1-2 mm [3]. Meanwhile, millimeter sized single crystals of $Ca(Fe_{1-x}Co_x)AsF$ with maximum $T_C$ = 21 K were grown also using the similar CaAs flux method [11]. As an alternative to CaAs flux method, $(Ca_{0.89}Na_{0.11})FeAsF$ up to 2×2 $mm^2$ in size with $T_C$ = 34.5 K were grown from NaCl flux cooling from 950 ℃ to room temperature [4]. With the commonly used CaAs or NaCl flux method, the crystal growth of CaFeAsF usually proceeds by a linear cooling from a soaking temperature to low temperature. Obviously, the small crystal size is attributed to spontaneous nucleation such that a number of additional crystals were produced during the growth process.

In order to increase the crystal size, seed crystals are necessary to prevent spontaneous nucleation in solutions. However, it is very difficult to use seeding technique in high-temperature solutions sealed in a quartz tube. As an alternative to seed crystals, oscillating temperature technique is a kind of temperature controlling method during crystal growth by employing some temperature interval cycles at the liquidus temperature followed by normal cooling of the flux or melt or by superposing the oscillation all over the normal cooling curve. Only the largest nuclei survive and grow while the smaller ones redissolve during the peak part of the temperature oscillation cycle. The reduction of the number of nuclei can be enhanced by the temperature oscillating technique which was proposed by Schäfer for chemical transport reactions, by Hintzmann and Müller-Vogt for high-temperature solutions and by Kolb and Laudise for hydrothermal flux growth [13-15]. Enlightened by the above pioneer work, the oscillating temperature technique was employed to grow large CaFeAsF single crystals, which reduces the number of CaFeAsF nuclei due to the temperature oscillation and enhances the growth of preferred nuclei depending on the temperature distribution in the crucible. Besides the temperature oscillating technique, other experimental conditions such as convection conditions of CaFeAsF solutes in the solution, the temperature gradient at the growth interface and the shape of crucible were also carefully considered for a stable and continuous growth process.

As an innovative application of the oscillating temperature technique, further details of our crystal growth experiments are given below. Ca granules (purity 99.5%, Alfa Aesar) and As powder (purity 99.995%, Alfa Aesar) with Ca : As = 1 : 1 in molar ratio were sealed in a quartz tube under vacuum and the quartz tube was placed in a box furnace which was heated to 700 °C for 12 hours. Then, the product was ground with an agate mortar and pestle, sealed in a quartz tube again. Repeat the above process several times to obtain a homogeneous CaAs powder. The CaAs powder (24.7112 g), $FeF_2$ (purity 99.5%, Alfa Aesar, 1.6803 g) and Fe powder (purity 99.98%, Alfa Aesar, 1.0000 g) were mixed thoroughly with the molar ratio CaAs : $FeF_2$ : Fe = 12 : 1 : 1 and pressed into separate columnar shapes to reduce the volume so as to be easily placed in a $Al_2O_3$ crucible with internal diameter of 12 mm and 15 cm in length. The crucible was sealed in a quartz tube with internal diameter of 20 mm and 20 cm in length under vacuum ($10^{-5}$ bar) and the quartz tube was placed in a vertical furnace with double temperature zones as shown in the schematic of Fig. 1(a). The double zones were heated to 1200 °C (bottom zone) and 1100 °C (upper zone) and soaked at this temperature for 24 h in order to ensure the dissolution of CaFeAsF solutes to the most extent. Then, for the purpose of reducing the number of nuclei two oscillating temperature procedures of bottom and upper temperature zones displayed in Fig. 1(b) were used during the whole growth process where it provides a slightly higher bottom temperature zone and a constant upward temperature gradient (~5 °C/cm). After the sufficient dissolution for CaFeAsF at

1200 ℃ the temperature (bottom zone) is lowered to 1100 ℃ at a rate of 2 ℃/h. Most of the crystals formed during this initial cooling are dissolved when the solution is heated to 1150 ℃ in 10 hours. This procedure is repeated until temperature reaches 950 ℃ such that only a few CaFeAsF single crystals have "survived". This constant temperature gradient (~5 ℃/cm) during the whole growth process helps to dissolve the CaFeAsF in molten CaAs flux at the bottom portion of the crucible and transport it upward by convection to crystallize in the cooler top portion of the solution by the chemical reaction $2CaAs + FeF_2 + Fe \rightarrow 2CaFeAsF$. Fig. 1(d) shows the flowchart of the oscillating temperature of the bottom zone during the whole growth process. In the end, by exposing the resultant to air in the fume hood for a few days, the flux CaAs decomposed and the shining CaFeAsF crystal plates were obtained.

The micro-morphology and chemical composition of the largest CaFeAsF single crystal were examined by scanning electron microscope (SEM) and energy dispersive spectrometer (EDS) on Phenom ProX. Powder and single crystal x-ray diffraction measurements were carried out at room temperature on an x-ray diffractometer (Rigaku UltimaIV) using Cu $K_\alpha$ radiation. Experiments of x-ray rocking curve were performed on a double-crystal diffractometer (Smartlab high resolution) equipped with a Ge (2 2 0) monochromator. The resistivity of crystal sample was measured on Quantum Design PPMS-9 using the standard 4-probe method.

3. **Results and discussions**

Fig. 1(c) shows the photos for some typical as-grown CaFeAsF single crystals and the largest as-grown CaFeAsF single crystal is nearly 8 mm in lateral size and 0.27 mm in thickness. This crystal size is far larger than previously reported in all dimensions. Fig. 2(a) shows the micro-morphology of the largest CaFeAsF single crystal taken by scanning electron microscope (SEM) from which the flat and tetragonal cleavage surface of (0 0 1) plane can be clearly seen. Displayed in Fig. 2(b) is the micro-morphology of lateral surface of CaFeAsF single crystal with layered texture. The thickness of the lateral surface is up to 270 μm. The chemical composition of the crystals was examined by the EDS analysis and one of the typical EDS spectrums taken on the rectangle area in Fig. 2(a) is shown in Fig. 2(c). The chemical composition is estimated to be Ca : Fe : As = 25.97 : 25.64 : 26.39 ≈ 1 : 0.99 : 1.01. The composition of the crystal is very homogeneous as demonstrated in the insets of fig. 2(c). Even the EDS is not sensitive to light element F, the rectangle mapping areas with blue, red, green and purple represent the homogeneous distribution of Ca, Fe, As as well as F respectively.

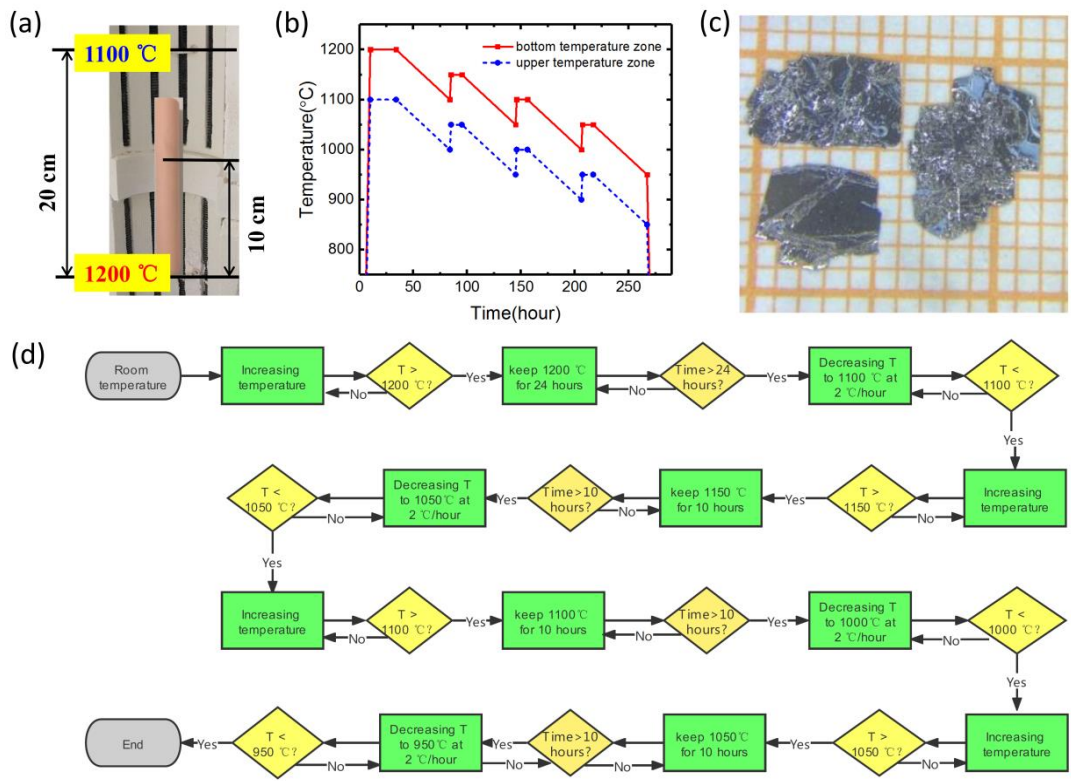

**Fig. 1.** (a) Experimental schemes for a convective CaAs solution from which the large CaFeAsF single crystals have been grown. (b) Temperature oscillating procedure in order to decrease the number of spontaneously nucleated crystals. (c) Photograph of typical CaFeAsF crystal pieces with typical dimensions about 6-8 mm grown in present work. The background mesh division is 1 mm. (d) The flowchart of the oscillating temperature at the bottom zone."

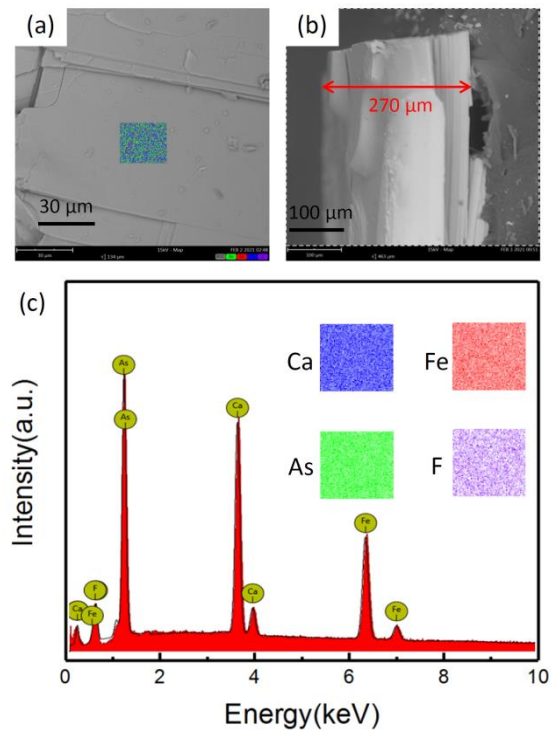

**Fig. 2.** (a-b) SEM image taken from the front and lateral surface of CaFeAsF crystal plate. (c) The EDS microanalysis spectrum taken on the rectangle area of CaFeAsF single crystal. The inset shows the elements distribution over the front surface of CaFeAsF single crystal.

Fig. 3 shows the results of CaFeAsF powder and single crystal x-ray diffraction (XRD) patterns at room temperature. Powder XRD demonstrates that all the main diffraction peaks of CaFeAsF can be well indexed with a previously reported tetragonal structure with the space group of P4/nmm, except for one weak peak from the $FeF_3$ impurity possibly due to the insufficient dissolution of Fe powder with the chemical reaction $3CaAs + 3FeF_2 + Fe \rightarrow 3CaFeAsF + FeF_3$. The lattice constants are calculated and compared with the data from ICDD (PDF No. 04-015-5127) and other report in Table 1. Only (0 0 l) reflections are observed from the single crystal XRD pattern, indicating that the single crystals are in perfect (0 0 1) orientation.

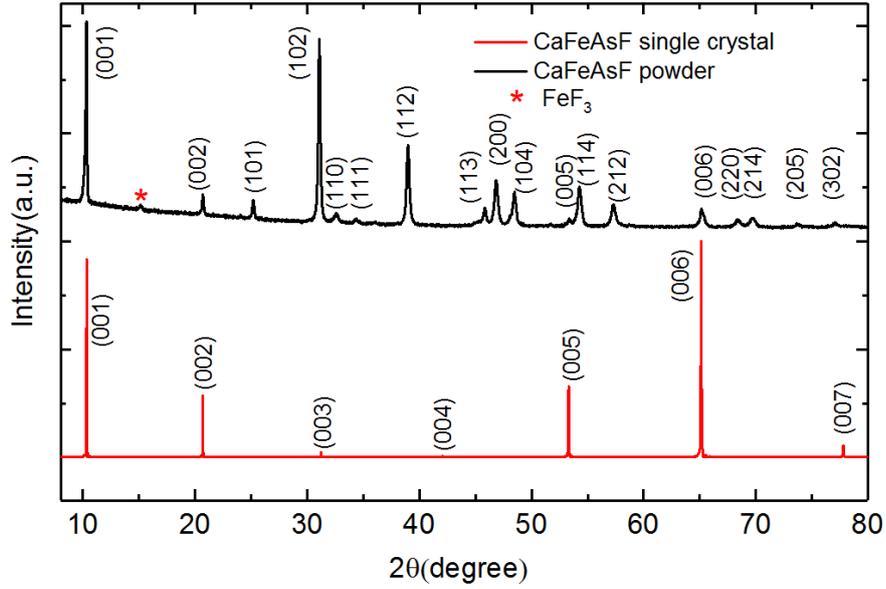

**Fig. 3.** Powder and single crystal x-ray diffraction patterns of CaFeAsF. All the main diffraction peaks from powder are indexed on a previously reported tetragonal structure, with lattice parameters being a = 3.8746(4) Å, c = 8.5824(12) Å, except for one weak peak from the $FeF_3$ impurity as indicated by asterisks. Only (0 0 l) reflections are observed from the typical XRD pattern obtained on the CaFeAsF single crystal indicating that the single crystals are in perfect (0 0 1) orientation.

**Table 1. Comparison of lattice constants of CaFeAsF**

| growth method | maximum lateral size | a (Å) | c (Å) | Ref. |
|---|---|---|---|---|
|  |  | 3.878 | 8.593 | ICDD |
| self-flux | 1-2 mm | 3.8774(4) | 8.5855(10) | [3] |
| oscillating temperature | 8 mm | 3.8746(4) | 8.5824(12) | This work |

The crystalline perfection of the CaFeAsF single crystal grown in this work was checked by double-crystal x-ray rocking curves of (0 0 5) Bragg reflection, as plotted in Fig. 4(a). The full-width-at-half maximum (FWHM), which is correlated with the crystal mosaicity, is as small as 0.163° indicating that the CaFeAsF crystals are of high crystalline quality. The inset of Fig. 4(a) shows an example of back-scattering x-ray Laue patterns obtained on (0 0 1) plane (*ab* plane) of CaFeAsF single crystal. The result shows very sharp and uniform diffraction spots with the 4-fold axis symmetry demonstrating no twin crystals inside the CaFeAsF single crystal. Temperature dependence of resistivity in the temperature range from 3 to 300 K for CaFeAsF single crystal under magnetic field from 0 T to 9 T is shown in Fig. 4(b). The transition temperatures, $T_s$ = 118 K and $T_{AFM}$ = 107 K, have been detected from the sharp change and clear kink in the resistivity data, which are slightly lower than previously reported [3]. While the resistivity above $T_{AFM}$ = 107 K remains nearly unchanged even under the magnetic field up to 9 T, it increases gradually with the increasing magnitude of magnetic field below $T_{AFM}$ which can be probably attributed to the magnetoresistance effect".

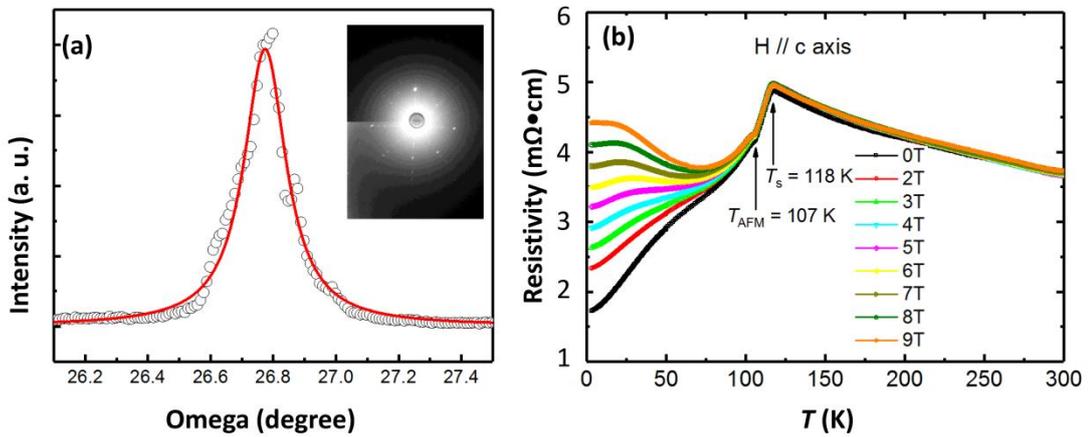

**Fig. 4.** (a) The inset shows x-ray rocking curve of the (0 0 5) Bragg reflection for a CaFeAsF single crystal taken by double-crystal x-ray diffraction using Cu $K_{\alpha1}$ radiation. The inset shows the Laue picture of CaFeAsF single crystal. (b) Temperature dependence of resistivity under different magnetic field for CaFeAsF single crystal measured in a wide temperature range from 3 K to 300 K."

4. **Conclusions**

In conclusion, the largest CaFeAsF single crystal with 8 mm in lateral size and 0.27 mm in thickness has been obtained by the successful application of the oscillating temperature technique. The crystal size is far larger than previously reported in all dimensions and the double-crystal x-ray rocking curve indicates high crystalline quality of CaFeAsF single crystals. Not only large CaFeAsF single crystals lay the foundation for the future neutron scattering experiments which will help to understand the mechanism of iron-based superconductivity but also shed light on other superconductors with small crystal size by further application of the oscillating temperature technique.

## Acknowledgments

The work was supported by the National Key Research and Development of China (Grant No. 2018YFA0704200), the National Natural Science Foundation of China (Grant No. 12004418) and the Strategic Priority Research Program of Chinese Academy of Sciences (Grant No. XDB25000000).

## References

[1] Y. Kamihara, T. Watanabe, M. Hirano and H. Hosono H, Iron-Based Layered Superconductor La[$O_{1-x}F_x$]FeAs (x = 0.05-0.12) with $T_c$ = 26 K, *J. Am. Chem. Soc.* **130** 3296 (2008). https://doi.org/10.1021/ja800073m.

[2] S. Matsuishi, Y. Inoue, T. Nomura, H. Yanagi, M. Hirano and H. Hosono, Superconductivity Induced by Co-Doping in Quaternary Fluoroarsenide CaFeAsF, *J. Am. Chem. Soc.* **130** 14428 (2008). https://doi.org/10.1021/ja806357j.

[3] Y. H. Ma, H. Zhang, B. Gao, K. K. Hu, Q. C. Ji, G. Mu, F. Q. Huang, X. M. Xie, Growth and characterization of millimeter sized single crystals of CaFeAsF, *Supercond. Sci. Technol.* **28** 085008 (2015). https://doi.org/10.1088/0953-2048/28/8/085008.

[4] L. Shlyk, K. K. Wolff, M. Bischoff, E. Rose, T. Schleid and R. Niewa, Crystal structure and superconducting properties of hole-doped $Ca_{0.89}Na_{0.11}$FFeAs single crystals, *Supercond. Sci. Technol.* **27** 044011 (2014). https://doi.org/10.1088/0953-2048/27/4/044011.

[5] H. Okada, H. Takahashi, S. Matsuishi, M. Hirano, H. Hosono, K. Matsubayashi, Y.

Uwatoko and H. Takahashi, Pressure dependence of the superconductor transition temperature of Ca(Fe$_{1-x}$Co$_x$)AsF compounds: A comparison with the effect of pressure on LaFeAsO$_{1-x}$F$_x$, *Phys. Rev. B* **81** 054507 (2010). https://doi.org/10.1103/PhysRevB.81.054507.

[6] P. Cheng, B. Shen, G. Mu, X. Zhu, F. Han, B. Zeng and H. H. Wen, High-$T_c$ superconductivity induced by doping rare-earth elements into CaFeAsF, *Europhys. Lett.* **85** 67003 (2009). https://doi.org/10.1209/0295-5075/85/67003.

[7] Z. A. Ren, W. Lu, J. Yang, W. Yi, X. L. Shen, Z. C. Li, G. C. Che, X. L. Dong, L. L. Sun, F. Zhou, Z. X. Zhao, Superconductivity at 55K in Iron-Based F-Doped Layered Quaternary Compound Sm[O$_{1-x}$F$_x$]FeAs, *Chin. Phys. Lett.* **25** 2215 (2008). https://doi.org/10.1088/0256-307X/25/6/080.

[8] T. Terashima, H. T. Hirose, D. Graf, Y. Ma, G. Mu, T. Hu, K. Suzuki, S. Uji, H. Ikeda, Fermi Surface with Dirac Fermions in CaFeAsF Determined via Quantum Oscillation Measurements, *Phys. Rev. X* **8** 011014 (2018). https://doi.org/10.1103/PhysRevX.8.011014.

[9] B. Xu B, H. Xiao, B. Gao, Y. H. Ma, G. Mu, P. Marsik, E. Sheveleva, F. Lyzwa, Y. M. Dai, R. P. S. Lobo, C. Bernhard, Optical study of Dirac fermions and related phonon anomalies in the antiferromagnetic compound CaFeAsF, *Phys. Rev. B* **97** 195110 (2018). https://doi.org/10.1103/PhysRevB.97.195110.

[10] J. Tao, S. Li, X. Y. Zhu, H. Yang and H. H. Wen, Growth and transport properties of CaFeAsF$_{1-x}$ single crystals, *Sci. China* **57** 632 (2014). https://doi.org/10.1007/s11433-014-5422-4.

[11] Y. H. Ma, K. K. Hu, Q. C. Ji, B. Gao, H. Zhang, G. Mu, F. Q. Huang and X. M. Xie, Growth and characterization of CaFe$_{1-x}$Co$_x$AsF single crystals by CaAs flux method, *J. Cryst. Growth* **451** 161 (2016). https://doi.org/10.1016/j.jcrysgro.2016.07.029.

[12] S. Price, Y. X. Su, Y. Xiao, D. T. Adroja, T. Guidi, R. Mittal, S. Nandi, S. Matsuishi, H. Hosono and T. Bruckel, Evidence of Spin Resonance Signal in Oxygen Free Superconducting CaFe$_{0.88}$Co$_{0.12}$AsF: An Inelastic Neutron Scattering Study, *J. Phys. Soc. Jpn.* **82** 104716 (2013). http://dx.doi.org/10.7566/JPSJ.82.104716.

[13] H. Schäfer, Chemical Vapor transport reactions, *Academic Press Inc.*, New York, 1964. https://doi.org/10.1016/C2013-0-12396-3.

[14] W. Hintzmann and G. Müller-Vogt, Crystal growth and lattice parameters of rare-earth doped yttrium phosphate, arsenate and vanadate prepared by the oscillating

temperature flux technique, *J. Cryst. Growth* **5** 274 (1969). https://doi.org/10.1016/0022-0248 (69) 90056-6.

[15] E. D. Kolb and R. A. Laudise, The solubility of trigonal Se in $Na_2S$ solutions and the hydrothermal growth of Se, *J. Cryst. Growth*, **8** 191 (1971). https://doi.org/10.1016/0022-0248(71)90141-2.